\UseRawInputEncoding
\documentclass[conference]{IEEEtran}

\usepackage{graphicx} 
\usepackage{amsmath}  
\usepackage[utf8]{inputenc} 
\usepackage[T1]{fontenc}    
\usepackage{url}      
\usepackage{hyperref} 
\usepackage{svg}
\usepackage{hyperref}
\usepackage{url}

\begin{document}

\title{Human Empathy as Encoder: AI-Assisted Depression Assessment in Special Education}

\author{
\IEEEauthorblockN{Boning Zhao}
\IEEEauthorblockA{\textit{Tandon School of Engineering} \\
\textit{New York University} \\
New York, USA \\
bz2518@nyu.edu}
\and

\IEEEauthorblockN{Xinnuo Li}
\IEEEauthorblockA{\textit{Tandon School of Engineering} \\
\textit{New York University} \\
New York, USA \\
xl5454@nyu.edu}
\and
\IEEEauthorblockN{Yutong Hu}
\IEEEauthorblockA{\textit{College of Arts \& Science} \\
\textit{New York University} \\
New York, USA \\
yh4872@nyu.edu}

}


\maketitle

\begin{abstract}
Assessing student depression in sensitive environments like special education is challenging. Standardized questionnaires may not fully reflect students' true situations. Furthermore, automated methods often falter with rich student narratives, lacking the crucial, individualized insights stemming from teachers' empathetic connections with students. Existing methods often fail to address this ambiguity or effectively integrate educator understanding. To address these limitations by fostering a synergistic human-AI collaboration, this paper introduces Human Empathy as Encoder (HEAE), a novel, human-centered AI framework for transparent and socially responsible depression severity assessment. Our approach uniquely integrates student narrative text with a teacher-derived, 9-dimensional "Empathy Vector" (EV), its dimensions guided by the PHQ-9 framework,to explicitly translate tacit empathetic insight into a structured AI input enhancing rather than replacing  human judgment. Rigorous experiments optimized the multimodal fusion, text representation, and classification architecture, achieving 82.05\% ± 0.58\% accuracy via 5-fold cross-validation for 7-level severity classification. This work demonstrates a path toward more responsible and ethical affective computing by structurally embedding human empathy.
\end{abstract}



\section{Introduction} 
The motivation for this research stems from firsthand observations in special education settings, where supporting student mental well-being, particularly regarding depression, is paramount. These environments, with their typically low student-to-teacher ratios, foster deep teacher familiarity with individual needs and contexts. However, depression assessment here presents unique challenges: standardized self-report questionnaires like PHQ-9 \cite{PlaceholderKeyQuestionnaireLimitations} often fail to capture students' authentic emotional states, while their natural conversations and writings can reveal richer insights. This creates a critical tension—educators possess invaluable empathy and contextual understanding from sustained interactions, yet without extensive psychological training, they struggle to systematically interpret vague narrative texts for signs of depression.

The challenge of interpreting student narratives is further compounded when considering purely automated approaches. Existing affective computing models often struggle with the nuances of expressed mental states like depression, potentially oversimplifying the assessment \cite{PlaceholderKeyAffectiveComputingLimitations}. Moreover, the growing adoption of large language models, especially cloud-based services, raises significant privacy concerns regarding sensitive student data \cite{PlaceholderKeyLLMPrivacyRisk}. Thus, the limitations of standalone human interpretation and the shortcomings of current AI tools collectively demand a new generation of assessment solutions grounded in social responsibility and synergistic human-AI collaboration \cite{PlaceholderKeyResponsibleAI}

To navigate these complex challenges, this paper introduces the "Human Empathy as Encoder (HEAE)" framework, a novel human-centered AI paradigm. HEAE fundamentally reorients depression assessment by structurally integrating teachers' tacit empathetic insights and experiential understanding directly into the AI analysis process. This approach complements and amplifies human judgment rather than attempting to replace it, fosters greater transparency compared to opaque automated systems and captures individualized nuances often missed by standardized questionnaires. Our work addresses implementation challenges through an innovative LLM-assisted annotation pipeline leveraging "golden seeds"—expert-annotated examples, to produce high-quality, nuanced training data for the HEAE model.
Architecturally, HEAE embodies the design principle of "structured simplicity + targeted complexity," which strategically allocates model sophistication where it most benefits human-AI collaboration. Beyond achieving 82.05\% ± 0.58\% accuracy via 5-fold cross-validation  accuracy in 7-level depression severity classification, this work provides both a privacy-preserving assessment tool for educational settings and a broader template for developing responsible, human-centered affective computing systems.
\section{Related Work} 

\subsection{Natural Language Processing for Depression Detection} 
The use of Natural Language Processing (NLP) to detect depression from text data has become a significant area of research. Researchers leverage techniques like sentiment analysis, linguistic feature extraction, and increasingly, deep learning models to analyze text from sources such as social media posts, web forums, and electronic health records \cite{NLPDepressionDetectionGeneral}. Deep learning architectures, including transformer-based models like RoBERTa\cite{liu2019roberta} and sequence models like BiLSTM\cite{graves2005framewise}, have shown considerable success in capturing linguistic cues associated with depression, often achieving high accuracy in distinguishing between depressed and non-depressed individuals \cite{DeepLearningDepressionSuccess}. However, significant challenges remain. While deep learning models are increasingly applied to classify depression severity or subtypes based on linguistic patterns, accurately capturing these fine-grained distinctions often proves difficult, especially when models rely heavily on surface-level features like keyword frequencies or sentiment scores \cite{Chancellor2020SurfaceLimitations}. Furthermore, these models frequently struggle with other nuanced emotional information, such as masked distress or ambiguous suicidal cues \cite{Zirikly2019CLPsych}. Ethical considerations regarding data privacy and bias in NLP models are also paramount \cite{NLPEthics}.

\subsection{Multimodal Approaches in Mental Health Assessment} 
Recognizing that mental health states manifest through various channels, multimodal approaches aim to achieve more comprehensive, robust, and accurate assessments by integrating complementary information from diverse sources like text, audio, and visual data. \cite{MultimodalReviewCohn, MultimodalReviewAudiovisual}.

However, many current multimodal approaches, particularly in high-context settings like special education, tend to rely heavily on sensor-derived or readily quantifiable data. This focus can lead to overlooking the rich background information and nuanced judgments possessed by human experts (such as teachers) who are familiar with the individuals \cite{AlSahili2024Survey}. 
Furthermore, even when diverse data streams are considered, including potentially qualitative human insights, Multimodal Machine Learning (MMML) techniques face ongoing challenges. Effectively aligning and fusing such heterogeneous data types, and developing models that can meaningfully interpret these complex combined signals, remain significant technical hurdles in the field \cite{MultimodalFusionChallenges}. 

\subsection{Human-in-the-Loop (HITL) and Human-AI Collaboration} 
In sensitive domains like healthcare and education, the concept of Human-in-the-Loop (HITL) AI and, more broadly, Human-AI Collaboration is gaining traction. These approaches intentionally incorporate human expertise into the AI workflow—which can involve humans providing labels, verifying AI predictions, or taking actions based on AI suggestions—with goals that often include improving accuracy, handling edge cases, ensuring alignment with human values, and maintaining human oversight over critical decisions \cite{Wu2022SurveyHITL, HITLDefinitionStanford, HACReviewHealth, HACReviewEducation}. However, while these general HITL and collaborative frameworks are broadly applied, effectively eliciting, representing, and incorporating deep, contextual, and empathetic understanding from domain experts remains a significant challenge, particularly for nuanced tasks such as interpreting student narratives in special education settings \cite{natarajan2024human}. This highlights an ongoing opportunity to develop more specialized mechanisms for such integration. Such human-centered collaborative systems contrast with AI designed to operate autonomously or replace human professionals entirely.

\subsection{Explainability, Ethics, and Responsibility in Affective Computing} 
As affective computing systems become more capable, ethical considerations—including data privacy, potential biases in emotion recognition across different demographics, the risk of emotional manipulation, and the need for user consent—become crucial \cite{AffectiveEthicsRG, AffectiveEthicsPDF}. The "black box" nature of many complex AI models raises concerns about transparency and accountability, which Explainable AI (XAI) seeks to address by making AI decision-making processes more understandable to humans \cite{ExplainableAIHealthcareDeeploy, ExplainableAIStressIJRIAS}.  Within mental health, principles of responsible AI emphasize human dignity, avoiding manipulation, preventing bias, ensuring human supervision over critical decisions, building trust, and showing empathy; consequently, there is a growing call for affective computing systems that are not only accurate but also transparent, fair, accountable, and ethically deployed \cite{ResponsibleAIEthicsSmythos, AffectiveEthicsPDF}. However, fully achieving these ideals and ensuring that AI systems, particularly in affective computing, are truly transparent and reliably align with human values and oversight remains an ongoing challenge.\cite{PlaceholderKeyResponsibleAI}

\subsection{Positioning the Current Work} 
Our work builds upon these areas while proposing a novel approach distinct from existing methods. While we leverage NLP and operate within a multimodal framework, our second modality is unique: the structured EV, which is derived from special education teachers applying their empathetic insights and contextual understanding to student narratives, using PHQ-9 dimensions as a structuring framework. This differs significantly from typical multimodal approaches that use raw audio/visual data \cite{MultimodalReviewCohn, MultimodalReviewAudiovisual}. Our HEAE framework represents a specific instantiation of Human-AI Collaboration \cite{HACReviewHealth}, going beyond standard HITL \cite{HITLDefinitionStanford} by structurally encoding teacher's empathetic insights a core input signal, designed explicitly to augment teacher insight. This design inherently promotes transparency by making the human contribution explicit, addressing ethical concerns regarding black-box AI \cite{ExplainableAIHealthcareDeeploy} in sensitive assessments, aligning with the objectives of developing reliable, safe, and trustworthy human-Centered AI \cite{ShneidermanHCAI}. By structurally embedding empathy into AI systems, our work not only demonstrates an effective assessment approach but also offers foundational insights for designing more socially responsible, transparent, and ultimately more trustworthy affective computing systems for real-world educational contexts.

\section{Methodology} 
This section details our HEAE framework, the dataset creation process, the optimized model architecture, and the experimental setup.

\subsection{Conceptual Framework: The Human Empathy as Encoder} 
The HEAE framework structurally encodes teachers' tacit empathetic insights and experiential understanding into the depression assessment process, addressing limitations of both standardized questionnaires and automated text analysis. In our approach, special education teachers analyze student narratives through the lens of PHQ-9's nine dimensions, converting their contextual understanding into a quantified EV. Unlike conventional PHQ-9 administration, this dimension-focused method accommodates the fragmentary nature of emotional signals in natural writing and student narratives, where not all depression indicators may be explicitly present. 

Teachers assign scores using a custom 0-5 severity scale (0=Not present, 5=Extreme impact) for each of these nine dimensions, rather than the standard 0-3 scale, providing the necessary sensitivity for our target 7-level classification\footnote{The seven levels are: No Depression, Minimal Depression, Mild Depression, Moderate Depression, Moderately Severe Depression, Severe Depression, and Severe Depression with High Suicide Risk.}. The resulting EV serves alongside the raw narrative as dual model input, creating an inherently transparent process that preserves individualized insights while making them algorithmically actionable.

\subsection{Dataset Generation and Features} 
To create a robust dataset for training our model, we developed a novel annotation pipeline using public narratives from social media.

\subsubsection{Golden Seed Development} 
To establish a high-quality foundation and encapsulate our target annotation logic, we created a set of initial examples termed "golden seeds." Raw narratives were sourced from Reddit forums, with all user-identifying information removed.

These golden seeds underwent manual annotation by experts with substantial experience in special education settings. This crucial step defined two key components.First, we established a machine-interpretable mapping between narrative content and scores across nine EV dimensions using the previously described 0-5 severity scale for each dimension, without calculating a total score. Second, these expert-generated EV scores were converted into a final 7-level depression severity classification. To ensure this EV-to-severity conversion method aligns with established understanding, its consistency with the PHQ-9's approach to categorizing severity was validated on all 200 golden seeds, yielding a Cohen's Kappa of 0.705, indicating substantial inter-rater agreement, and an overall accuracy of 75.5\%. Each golden seed subsequently included the narrative text, its expert-generated 9-dimensional EV, the resulting final 7-level severity label, and detailed explanations justifying the EV scores, thus encapsulating the target reasoning process.

\subsubsection{Novel Automated Annotation Pipeline} 
\begin{figure}[htbp]
    \centering
    \includegraphics[width=\linewidth]{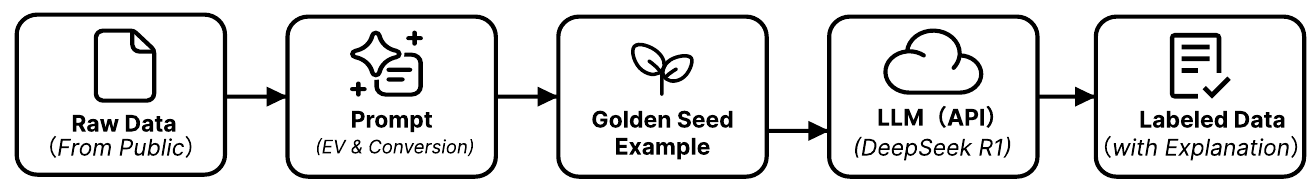} 
    \caption{Automated annotation pipeline using golden seeds and an LLM.}
    \label{fig:pipeline}
\end{figure}

To efficiently scale our dataset while maintaining consistent interpretation patterns, we developed an LLM-assisted annotation pipeline (Figure~\ref{fig:pipeline}). From the full set of 200 validated golden seeds, we carefully selected 20 diverse examples to serve as few-shot demonstrations for the LLM. This number was chosen to provide a rich variety of narrative styles and severity levels, while respecting the input token limitations of the language model. To further guide the LLM and ensure annotation consistency, the prompt was supplemented with explicit details regarding the Empathy Vector (EV) structure and our validated EV-to-severity conversion method, alongside instructions for the LLM to adopt a persona reflecting expertise in empathetic mental well-being assessment in educational contexts.

\subsubsection{Dataset Finalization and Characteristics} 
\label{sec:III.B.3} 
This annotation pipeline initially yielded approximately 30,000 labeled samples. After balancing various levels of depression to address skewness often found in online data which commonly overrepresented severe cases, making intermediate levels like moderate depression relatively harder to source sufficiently, we obtained approximately 17,500 samples for model development.

\subsubsection{Model Inputs, Target} 
\label{sec:Model input}
The model utilizes two primary inputs: (i)  Narrative Text, and (ii) EV (9-dim, 0-5 scale from teacher interpretation). The classification target is the 7-level depression severity assigned during data generation. 

\subsection{Final Model Architecture} 
\begin{figure*}[htbp] 
    \centering
    \includegraphics[width=\textwidth]{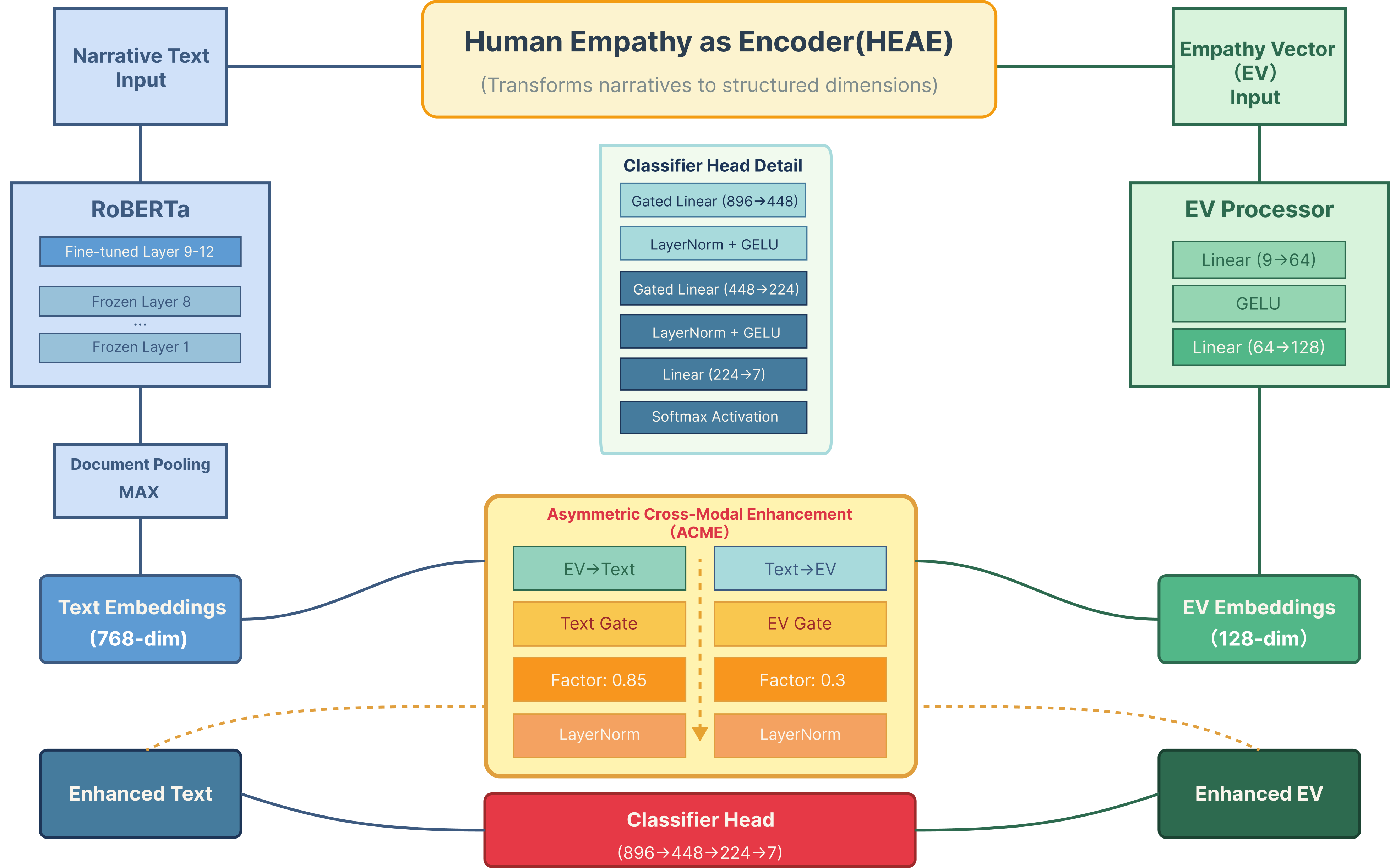}
    \caption{The architecture of the proposed Human Empathy as Encoder (HEAE) model.}
    \label{fig:architecture}
\end{figure*}

The final model architecture (Figure~\ref{fig:architecture}) integrates the text and EV modalities using components optimized through experiments detailed in Section~\ref{sec:experiments}. 

\subsubsection{Text Encoder} 
We utilize a pre-trained RoBERTa model.  The bottom 8 layers were frozen, while the top 4 layers were fine-tuned.

\subsubsection{Text Representation} 
A sliding window (512 token chunks, 256 stride) handles long narratives. Chunk embeddings are aggregated into a single document representation using Max Pooling, which effectively captures peak emotional signals, followed by Layer Normalization.

\subsubsection{Empathy Vector Processing} 
\label{sec:ev_processing}
The 9-dimensional EV is projected into a higher-dimensional space suitable for fusion using two linear layers (transforming dimensions 9 → 64 → 128) with GELU activation.

\subsubsection{Multimodal Fusion} 
\label{sec:fusion}
Our Asymmetric Cross-Modal Enhancement (ACME) mechanism is designed to effectively fuse the narrative text and EV modalities. It employs a controlled, gated interaction with fixed asymmetric factors to guide the enhancement process, prioritizing the influence of the EV on text representations.

Let $X_t$ be the processed text embedding and $X_e$ be the processed EV embedding. To dynamically control the information flow between these modalities, gate signals $G_t$ (for text) and $G_e$ (for EV) are first computed from the concatenated embeddings $[X_t; X_e]$ using separate linear layers followed by sigmoid activations. These gates modulate the subsequent cross-modal influence. The core asymmetric enhancement is then defined as:
\begin{align} 
    X_t^{enh} &= X_t + G_t \odot (W_{e \rightarrow t} X_e) \cdot \alpha_{e \rightarrow t} \label{eq:acme_text} \\
    X_e^{enh} &= X_e + G_e \odot (W_{t \rightarrow e} X_t) \cdot \alpha_{t \rightarrow e} \label{eq:acme_ev}
\end{align}
where $X_t^{enh}$ and $X_e^{enh}$ are the enhanced text and EV embeddings, respectively. $W_{e \rightarrow t}$ and $W_{t \rightarrow e}$ represent the weights of learnable linear transformations that map embeddings from one modality's space to the other. The symbol $\odot$ denotes element-wise multiplication. The fixed asymmetric enhancement factors are $\alpha_{e \rightarrow t} = 0.85$ (EV to Text) and $\alpha_{t \rightarrow e} = 0.30$ (Text to EV), reflecting a stronger guidance from the EV. 

The resulting enhanced embeddings, $X_t^{enh}$ and $X_e^{enh}$, are then individually passed through Layer Normalization. This entire controlled interaction, stabilized by Layer Normalization, proved superior to complex attention mechanisms in our experiments (see Section~\ref{sec:Optimizing Multimodal Fusion}). 

\subsubsection{Classifier Head} 
The concatenated features from enhanced text (768D) and EV (128D) representations are processed by a deep classifier head designed for effective feature transformation before final classification. This architecture employs sequential Gated Linear Units (GLUs)\cite{dauphin2017language} with GELU activation, progressively reducing dimensionality from the initial combined 896D (768D + 128D) to 448D and finally to 224D. Each GLU layer is followed by Layer Normalization with another GELU activation. A final linear layer maps the 224-dimensional features to the 7 output classes, followed by Softmax activation. This strategic depth proved essential for the fine-grained classification task.

\subsection{Experimental Setup and Training} 
\label{sec:III.D} 
Our curated dataset (detailed in Section~\ref{sec:III.B.3}) was evaluated using 5-fold cross-validation to ensure robust performance assessment. We employed the AdamW optimizer with a learning rate of 1e-5 and a batch size of 8. Model performance was evaluated using accuracy and macro F1-score across all folds.

\section{Experiments and Results} 
\label{sec:experiments} 
This section presents the empirical results of systematically optimizing the multimodal architecture for the 7-level depression severity classification task. 
\subsection{Baseline Model Performance} 
We first established a baseline model using RoBERTa text embeddings and processed EV embeddings, combined via simple feature concatenation followed by a basic classifier head.
This baseline achieved an accuracy of approximately 73\%, serving as the starting point for evaluating more sophisticated integration strategies.

\subsection{Optimizing Multimodal Fusion} 
We then investigated methods to improve the fusion of text and EV embeddings:
\label{sec:Optimizing Multimodal Fusion}
\subsubsection{Evaluating Fusion Strategies}
\begin{figure}[htbp]
    \centering
    \includegraphics[width=\linewidth]{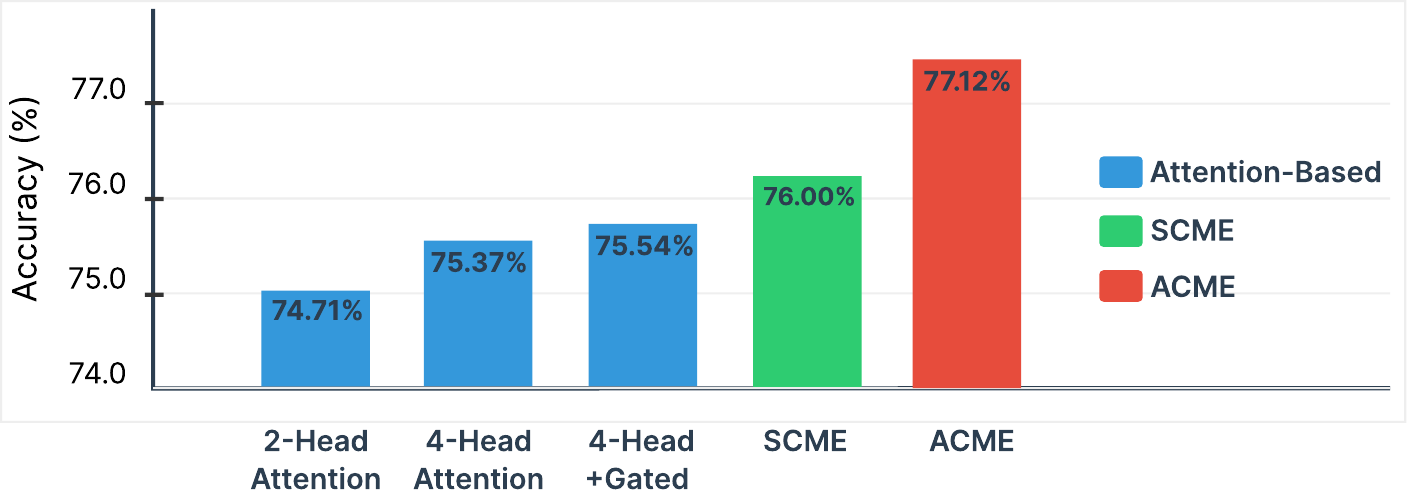} 
    \caption{Performance comparison of different multimodal fusion strategies.} 
    \label{fig:fusion_comparison}
\end{figure}

To effectively integrate the text and EV modalities, we first investigated the performance of standard multi-head cross-attention mechanisms, a common approach for multimodal fusion. As shown in Figure~\ref{fig:fusion_comparison}, even the most effective configuration of these attention models offered only a slight improvement over the simple concatenation baseline. This modest outcome motivated our exploration of more tailored fusion strategies. Consequently, we developed and evaluated our advanced fusion mechanisms: Symmetric Cross-Modal Enhancement (SCME) and Asymmetric Cross-Modal Enhancement (ACME). The design of the ACME mechanism is detailed in Section~\ref{sec:fusion}. In stark contrast, while SCME demonstrated further improvement, our ACME approach achieved significantly superior performance, reaching 77.12\% accuracy. This result clearly outperformed both the baseline and all tested attention-based models, thus validating this new direction.

\subsubsection{Optimizing Asymmetric Cross-Modal Enhancement} 
\label{sec:multimodal_fusion} 
Having established the superior potential of ACME over both complex attention and SCME , we further analyzed its optimization. Initial tests evaluating SCME found that the best-performing symmetric factor was 0.15, yielding 76.00\% accuracy. However, subsequent experiments exploring dynamically learned factors for the enhancement mechanism indicated a natural asymmetry was beneficial. This suggested that optimally leveraging the teacher-derived EV required careful, asymmetric calibration to provide essential grounding for interpreting unstructured text narratives, a core principle of our approach.

Based on this observation, we tested several fixed asymmetric factor configurations to identify the optimal setup. As shown in Table I, the configuration with $\alpha_{e \rightarrow t} = 0.85$ and $\alpha_{t \rightarrow e} = 0.30$ achieved the best performance at 77.12\% accuracy, outperforming both other asymmetric variants and dynamic factor learning.

\begin{table}[htbp]
\centering
\caption{Performance of ACME with Different Asymmetric Factor Configurations.}
\label{tab:acme_optimization}
\begin{tabular}{lcc}
\hline
\textbf{Configuration} & \textbf{Factors ($\alpha_{e \rightarrow t}$ / $\alpha_{t \rightarrow e}$)} & \textbf{Accuracy (\%)} \\
\hline
Asymmetric Variant 1 & 0.90 / 0.25 & 76.54 \\
\textbf{ACME (Optimal)} & \textbf{0.85 / 0.30} & \textbf{77.12} \\
Asymmetric Variant 2 & 0.80 / 0.35 & 76.16 \\
Dynamic Factor & Learned & 76.79 \\
\hline
\end{tabular}
\end{table}

This result empirically confirmed the value of stronger, asymmetrically weighted guidance from the  teacher's structured and empathetic judgment  provided by the EV.

This optimized ACME mechanism significantly outperformed the more complex multi-head cross-attention mechanisms (Sec IV.B.1, Figure~\ref{fig:fusion_comparison}). We hypothesize that complex attention struggles in this context due to potential challenges including:
\begin{itemize}
    \item The significant dimensionality mismatch between high-dimensional text embeddings (768D from RoBERTa) and the lower-dimensional processed EV (128D, see Sec~\ref{sec:ev_processing}), hindering effective alignment;
    \item The potential risk of diluting the sparse but critical information within the EV;
    \item The moderate dataset size (~17.5k samples), which may be insufficient to robustly train complex attention for this specific cross-modal interaction.
\end{itemize}
In contrast, the optimized ACME mechanism provided controlled interaction, effectively preserving teacher's empathetic insights while appropriately guiding the model's interpretation of emotional narratives. 

This optimized ACME configuration, demonstrating the benefits of calibrated human insight integration and strategic simplicity in fusion, was used in all subsequent experiments.

\subsection{Text Representation Optimization}
\label{sec:text_representation}

Following the optimization of our cross-modal fusion mechanism (Sec~\ref{sec:multimodal_fusion}), we next refined the strategy for representing long student narratives. Our initial approach involved segmenting narratives using a sliding window (512 tokens, 256 stride) to produce chunk embeddings.

We first explored methods to explicitly leverage inter-chunk information. However, experiments revealed that incorporating explicit chunk order via positional embeddings significantly reduced accuracy, and employing a more complex multi-head Transformer-based chunk aggregator failed to yield improvements over simpler aggregation of individual chunk CLS embeddings. These findings suggested that for our task, capturing the most salient emotional signals within each chunk and relying on the implicit context from a substantial token overlap (256-token stride proved optimal) was more effective than explicitly modeling precise chunk order or complex sequential dependencies between chunks. We therefore proceeded without explicit positional embeddings and focused on optimizing the pooling of chunk CLS embeddings.

\begin{figure}[htbp]
\centering
\includegraphics[width=0.8\linewidth]{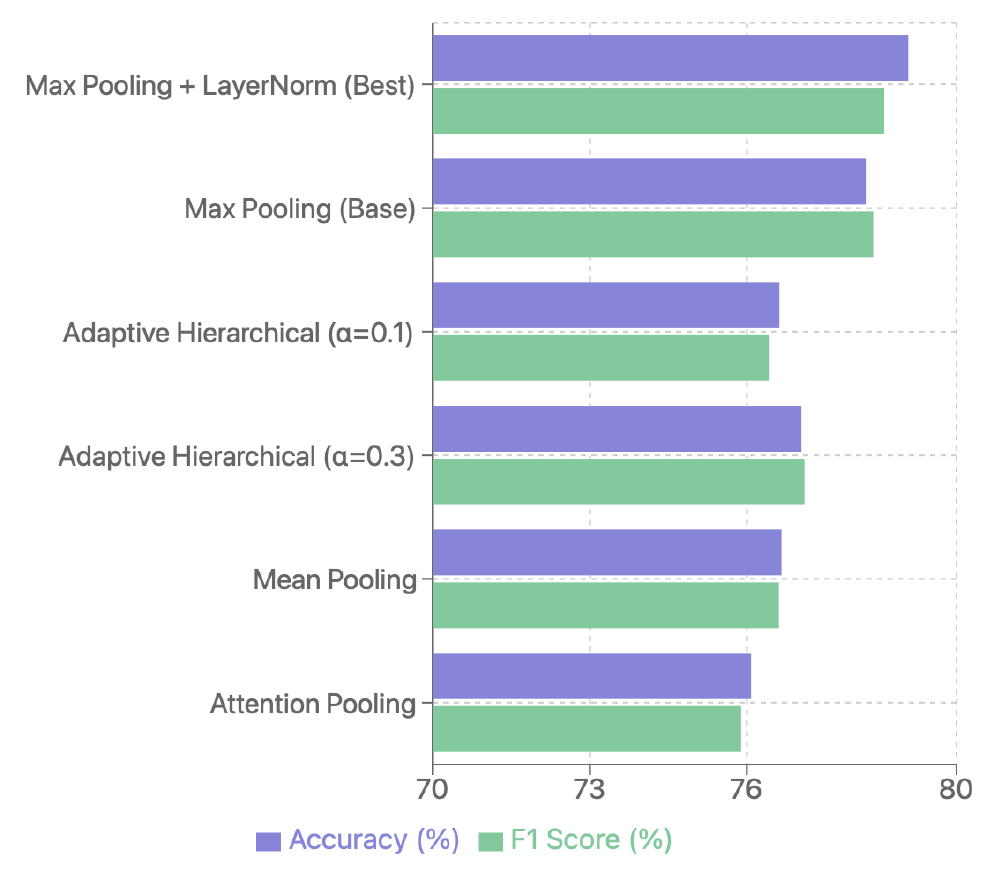}
\caption{Comparison of different pooling strategies for aggregating text chunk embeddings.}
\label{fig:pooling_comparison}
\end{figure}

Evaluating standard document pooling strategies (Figure~\ref{fig:pooling_comparison}), Max Pooling demonstrated clearly superior performance over alternatives like Mean Pooling, Attention Pooling, and Adaptive Hierarchical pooling. This highlighted the importance of capturing peak emotional signals from the narrative texts for assessing depression severity. Further significant improvement was achieved by incorporating Layer Normalization after Max Pooling, yielding an accuracy of 79.08\% and establishing 'Max Pooling + LayerNorm' as our optimal text representation strategy.

Importantly, while Max Pooling identifies these "peak signals," these are not interpreted naively (e.g., based solely on keywords such as "I want to die") within the HEAE framework. Instead, the ACME mechanism (Sec~\ref{sec:multimodal_fusion}) leverages the teacher-derived EV to provide essential context. This effectively grounds and validates text-based signals against the teacher's holistic, empathetic assessment, underscoring the indispensable role of structured human empathy in achieving nuanced and reliable judgment from student narratives.

\subsection{Classifier Head Optimization} 
\label{sec:classifier_optimization} 
\begin{figure}[htbp]
    \centering
    \includegraphics[width=\linewidth]{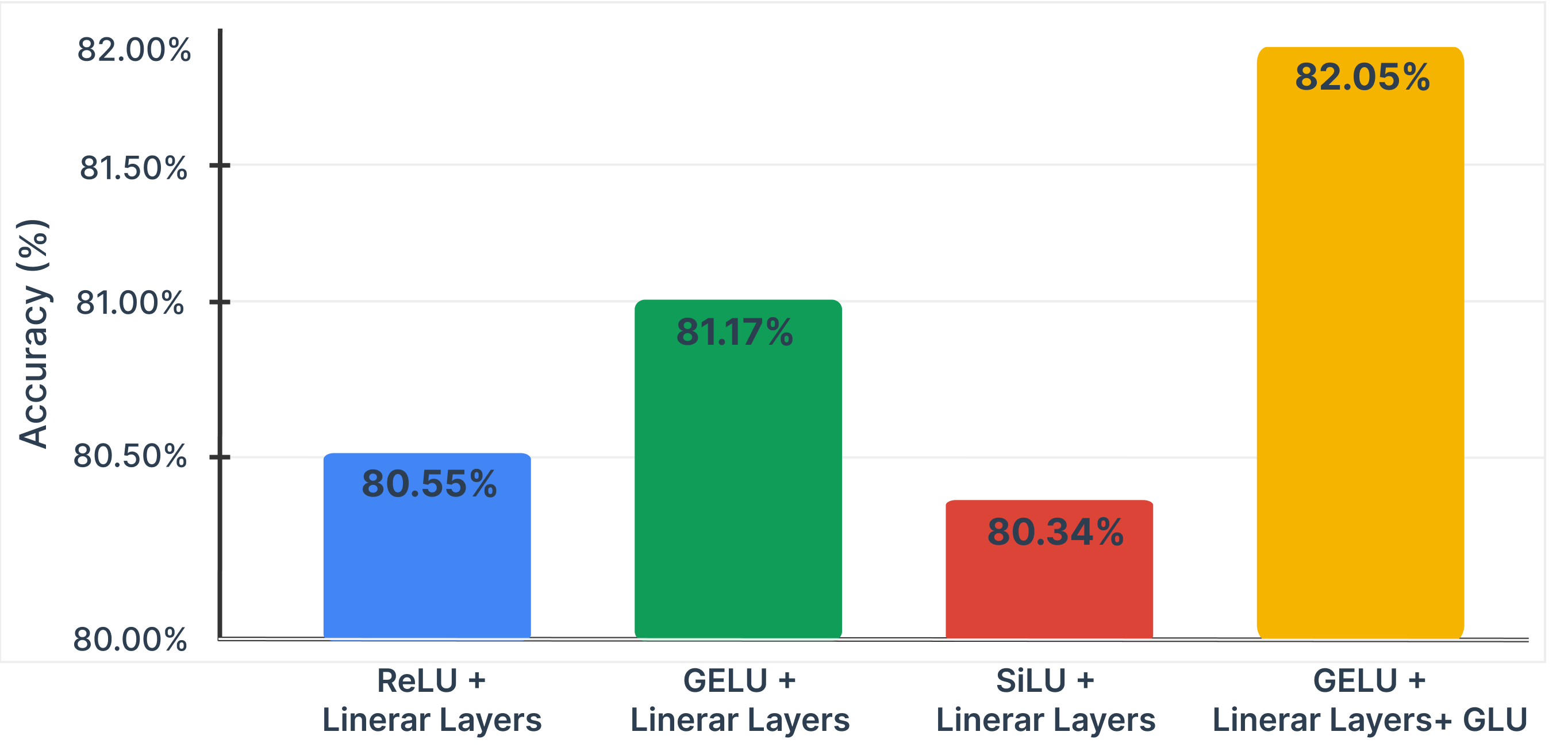} 
    \caption{Performance comparison of different classifier head architectures.}
    \label{fig:activation_comparison}
\end{figure}

With optimized fusion (Sec~\ref{sec:multimodal_fusion}) and text representation (Sec~\ref{sec:text_representation}), we refined the classifier head to better handle the fine-grained 7-level severity classification. We found that increasing the classifier depth by implementing a multi-layer architecture ((896D → 448D → 224D → 7 classes) significantly improved performance. Within this deeper structure, GELU activation yielded the best performance, outperforming ReLU and SiLU. Further enhancement was achieved by incorporating GLUs into this GELU-activated structure, bringing the final accuracy to 82.05\% (Figure~\ref{fig:activation_comparison}). 

This classifier's improved performance highlights a key architectural principle: unlike earlier model stages (e.g., fusion, document pooling) where tailored simpler mechanisms were optimal for preserving key signals, the final classification of rich, aggregated multimodal features demands significantly increased representational capacity. The deeper, gated architecture, incorporating GELU activation and GLUs, provided this necessary capacity, enabling effective learning of complex non-linearities and adaptive features to discern subtle differences among the 7 severity levels from the consolidated inputs.

\subsection{Final Model Performance} 
The final optimized HEAE model, incorporating all optimal components (asymmetric fusion, max pooling, and deep GELU+GLUs classifier head), achieved 82.05\% ± 0.58\% accuracy and 82.08\% macro F1-score for our 7-level depression severity classification task. This represents a significant improvement of approximately 9\% points over the baseline model (73\% accuracy), demonstrating the effectiveness of our human-empathy guided approach.

\section{Discussion} 
\subsection{Responsible Design Principles for Human-AI Collaboration}
The HEAE framework introduces a new paradigm for depression assessment in special education. By structuring teacher empathy into the EV as an explicit input, it achieves an organic fusion of human judgment and AI analysis. This method makes teachers' tacit professional insight explicit. It preserves the primacy of human judgment while enabling teachers, through AI assistance, to more effectively apply their rich empathetic understanding. This leads to assessments of student depression that are more nuanced and contextualized than what traditional standardized questionnaires typically allow. Our framework rejects the "AI replaces human" approach. Instead, it builds a complementary relationship, positioning technology as a means to enhance human empathy, not substitute it. More importantly, this reorientation creates an inherently more transparent and ethically sound evaluation system. It offers a blueprint for sensitive domains where maintaining core human judgment is essential.

Our experiments reveal a key design principle for human-AI collaboration: complexity should be strategically allocated, not uniformly applied. The tailored simplicity in modal fusion and representation extraction contrasts sharply with the targeted depth in classification. This reflects more than just technical choices; it embodies a philosophy regarding the transparency of the human-AI interaction. This "structured simplicity + targeted complexity" approach challenges the "more complex is better" tendency often seen in the pursuit of model performance. It suggests that truly effective human-AI systems should respect and leverage the respective strengths of both humans and AI. Therefore, we advocate prioritizing understandability and signal fidelity during early-stage information integration. At the same time, necessary computational sophistication should be permitted for the final, fine-grained decision-making stage. This balancing principle extends beyond specific algorithm choices and aligns with current ethical demands for AI transparency and explainability. It charts a new course for designing future responsible affective computing systems—ones that are both effective and ethically considerate, especially in highly sensitive areas like special education

\subsection{Implications for Socially Responsible and Explainable AI} 
With its ability to structurally integrate nuanced teacher empathy via the EV and student narratives, the HEAE framework offers significant implications for responsible, explainable, and privacy-preserving affective AI systems.

At its core, HEAE achieves transparency through its fundamental structure rather than relying on after-the-fact explanations. The EV creates a clear connection between teacher judgment and AI analysis, making the influence pathway visible from the outset.
This design approach offers inherent clarity that conventional black-box models lack. The explicit nature of this structure makes both the AI's reasoning process more traceable and the human input more examinable—qualities essential for identifying potential biases and building trust in the human-AI collaborative assessment process.

HEAE also embodies privacy-by-design principles essential for sensitive educational contexts. Its efficient architecture enables local processing, minimizing the need to transmit sensitive student data to external servers. This approach not only protects privacy but also makes sophisticated assessment technology accessible to resource-limited special education environments, demonstrating that responsible AI development can balance advanced capabilities with ethical deployment constraints.

By maintaining educators as integral participants, HEAE preserves human agency and accountability in high-stakes assessment. The system functions as an extension of professional judgment rather than a replacement, creating a collaborative framework where technology enhances rather than diminishes human expertise. This partnership model ensures that emotional intelligence in the assessment process stems from the synergy of human empathy and computational analysis, not from AI in isolation.

\section{Limitations and Future Work}
While HEAE demonstrates promise for human-AI collaborative depression assessment, several important limitations and future directions warrant consideration:

\textbf{Empathy Vector Subjectivity and Nuance:} The EV effectively structures teacher empathy, yet fully capturing the richness and variability of empathetic judgment remains challenging. The inherent subjectivity across different educators and student contexts suggests the need for research into consensus mechanisms among multiple raters and context-sensitive EV refinements, while preserving authentic human insight.

\textbf{Real-World Generalizability and Implementation:} Our current validation, while encouraging, occurred within a controlled experimental environment. Real-world special education settings present additional challenges including diverse narrative styles, varying teacher technological familiarity, and integration with existing workflows. Future deployment studies in authentic educational environments are essential to assess practical usability, effectiveness, and impact on teacher workload.

\textbf{Privacy and Ethical Deployment Considerations:} While our design prioritizes privacy through local processing, practical deployment requires further attention to data governance in educational settings. Future work should explore privacy-preserving techniques that enable model improvement without centralized data collection, and develop ethical protocols appropriate for vulnerable student populations and sensitive mental health data. These considerations are especially important given the heightened privacy requirements in special education contexts.

\section{Conclusion} 
This paper introduced the HEAE framework, demonstrating a novel human-AI collaborative paradigm for depression severity assessment in special education by structurally integrating teacher empathy. Beyond achieving effective classification performance, this approach offers a blueprint for developing more transparent, ethically sound, and human-centered affective computing systems in sensitive domains.

\section*{Data Availability Statement}
The "Golden Seeds" dataset, validation code 
are available on \url{https://huggingface.co/datasets/Plum551/golden_seeds_ev_depression}
\section*{Acknowledgments}  
This work is dedicated to the one whose profound inspiration and unwavering belief have reshaped my life -a soul whose quiet presence still guides this research.

\section*{ Ethical Impact Statement } 

This statement outlines the measures undertaken to ensure responsible data handling and alignment with ethical AI principles in the development of the Human Empathy as Encoder (HEAE) framework.

\textbf{Data Source} 
The dataset consists of publicly accessible posts from online mental health communities, with typical demographic representation of young adults. All data were sourced from public domains, with no data from minors or private sources collected. The dataset is used strictly for non-commercial research purposes, specifically advancing AI-driven tools for understanding and assessing mental well-being within educational contexts.

\textbf{Data Privacy} 
Recognizing the highly sensitive nature of the content—including self-disclosures of trauma, suicidality, and other deeply personal experiences—we implemented a comprehensive, multi-stage anonymization protocol:
\begin{itemize}
    \item Removal of direct Personally Identifiable Information (PII) using Named Entity Recognition (NER) and rule-based techniques.
    \item Generalization or removal of potentially identifying details such as precise ages, institutional names, exact geographical locations, and specific familial relationships.
    \item Removal of date-specific information and unique circumstantial identifiers that could contribute to re-identification.
    \item Additional NER tools with manual review to ensure thoroughness where automated methods might prove insufficient.
\end{itemize}
We acknowledge the inherent challenges in achieving complete anonymization when dealing with rich, personal narratives. To mitigate residual risk, the dataset is not publicly released and is maintained securely by the researcher for this research study only. This practice adheres to ethical principles of minimizing potential harm, maximizing research beneficence, and upholding respect for individual privacy and dignity.



\begin{thebibliography}{10}
\providecommand{\url}[1]{#1}
\csname url@samestyle\endcsname
\providecommand{\newblock}{\relax}
\providecommand{\bibinfo}[2]{#2}
\providecommand{\BIBentrySTDinterwordspacing}{\spaceskip=0pt\relax}
\providecommand{\BIBentryALTinterwordstretchfactor}{4}
\providecommand{\BIBentryALTinterwordspacing}{\spaceskip=\fontdimen2\font plus
\BIBentryALTinterwordstretchfactor\fontdimen3\font minus \fontdimen4\font\relax}
\providecommand{\BIBforeignlanguage}[2]{{%
\expandafter\ifx\csname l@#1\endcsname\relax
\typeout{** WARNING: IEEEtran.bst: No hyphenation pattern has been}%
\typeout{** loaded for the language `#1'. Using the pattern for}%
\typeout{** the default language instead.}%
\else
\language=\csname l@#1\endcsname
\fi
#2}}
\providecommand{\BIBdecl}{\relax}
\BIBdecl

\bibitem{PlaceholderKeyQuestionnaireLimitations}
\BIBentryALTinterwordspacing
K.~S. Button, D.~Kounali, L.~Thomas, N.~J. Wiles, T.~J. Peters, and G.~Lewis, ``Limitations of the patient health questionnaire ({PHQ-9}) in identifying anxiety and depression: many cases are undetected,'' \emph{British Journal of General Practice}, vol.~64, no. 619, pp. e130--e136, Feb 2014. [Online]. Available: \url{https://pmc.ncbi.nlm.nih.gov/articles/PMC3899353/}
\BIBentrySTDinterwordspacing

\bibitem{PlaceholderKeyAffectiveComputingLimitations}
\BIBentryALTinterwordspacing
S.~D. Kapadia, K.~Saha, and P.~Bhattacharya, ``Machine learning approaches for mental illness detection on social media: A systematic review of biases and methodological challenges,'' arXiv preprint arXiv:2410.16204v3, 2024, version 3. [Online]. Available: \url{https://arxiv.org/html/2410.16204v3}
\BIBentrySTDinterwordspacing

\bibitem{PlaceholderKeyLLMPrivacyRisk}
\BIBentryALTinterwordspacing
H.~Harvey and A.~Ashok, ``Implementing large language models in healthcare while balancing control, collaboration, costs and security,'' \emph{BMJ Health \& Care Informatics}, vol.~31, no.~1, p. e000957, Mar 2024. [Online]. Available: \url{https://pmc.ncbi.nlm.nih.gov/articles/PMC11885444/}
\BIBentrySTDinterwordspacing

\bibitem{PlaceholderKeyResponsibleAI}
\BIBentryALTinterwordspacing
G.~Sowemimo-Coker, J.~A. Robles-Zurita, R.~Ryan, and I.~Tachtsidis, ``Responsible ai integration in mental health research: Issues, guidelines, and best practices,'' \emph{JMIR Mental Health}, vol.~11, p. e55009, Jun 2024. [Online]. Available: \url{https://pmc.ncbi.nlm.nih.gov/articles/PMC11624515/}
\BIBentrySTDinterwordspacing

\bibitem{NLPDepressionDetectionGeneral}
\BIBentryALTinterwordspacing
S.~El~Idrissi, B.~Ben~Amor, and N.~Abid, ``Depression detection from social media textual data using natural language processing and machine learning techniques,'' in \emph{2024 International Conference on Digital Technologies and Applications (ICDTA)}, 2024, pp. 1--6. [Online]. Available: \url{https://ieeexplore.ieee.org/abstract/document/10441612}
\BIBentrySTDinterwordspacing

\bibitem{liu2019roberta}
Y.~Liu, M.~Ott, N.~Goyal, J.~Du, M.~Joshi, D.~Chen, O.~Levy, M.~Lewis, L.~Zettlemoyer, and V.~Stoyanov, ``Roberta: A robustly optimized bert pretraining approach,'' \emph{arXiv preprint arXiv:1907.11692}, 2019.

\bibitem{graves2005framewise}
A.~Graves and J.~Schmidhuber, ``Framewise phoneme classification with bidirectional lstm and other neural network architectures,'' \emph{Neural networks}, vol.~18, no. 5-6, pp. 602--610, 2005.

\bibitem{DeepLearningDepressionSuccess}
\BIBentryALTinterwordspacing
M.~J.~H. Bhuiyan, S.~B. Akintoye, E.~Deniz, R.~Mathew, M.~R. Islam, and M.~R. Nasim, ``Deep learning-based detection of depression and suicidal tendencies in social media data with feature selection,'' \emph{Diagnostics (Basel)}, vol.~14, no.~8, p. 795, Apr 2024. [Online]. Available: \url{https://pmc.ncbi.nlm.nih.gov/articles/PMC11939175/}
\BIBentrySTDinterwordspacing

\bibitem{Chancellor2020SurfaceLimitations}
S.~Chancellor and M.~De~Choudhury, ``Methods in predictive techniques for mental health status on social media: a critical review,'' \emph{NPJ Digital Medicine}, vol.~3, no.~1, pp. 1--11, 2020.

\bibitem{Zirikly2019CLPsych}
A.~Zirikly, P.~Resnik, {\"O}.~Uzuner, and K.~Hollingshead, ``Clpsych 2019 shared task: Predicting the degree of suicide risk in reddit posts,'' in \emph{Proceedings of the Sixth Workshop on Computational Linguistics and Clinical Psychology}, 2019, pp. 24--33.

\bibitem{NLPEthics}
\BIBentryALTinterwordspacing
S.~Almaiman, ``Screening for depression using natural language processing: Literature review,'' \emph{Journal of Medical Internet Research}, vol.~26, p. e55067, May 2024. [Online]. Available: \url{https://www.i-jmr.org/2024/1/e55067}
\BIBentrySTDinterwordspacing

\bibitem{MultimodalReviewCohn}
\BIBentryALTinterwordspacing
J.~F. Cohn, N.~Cummins, J.~Epps, R.~Goecke, J.~Joshi, and S.~Scherer, ``Multimodal assessment of depression from behavioral signals,'' \emph{Chapter in Computational Interaction, Oxford University Press}, 2019, accessed from author's website, chapter contribution. [Online]. Available: \url{https://www.jeffcohn.net/wp-content/uploads/2019/03/VII3-MultimodalAssessmentDepression_Final_V05.pdf}
\BIBentrySTDinterwordspacing

\bibitem{MultimodalReviewAudiovisual}
\BIBentryALTinterwordspacing
M.~R. Islam, M.~S. Islam, M.~M. Rahman, and M.~Hossain, ``Deep learning for depression recognition with audiovisual cues: A review,'' \emph{Knowledge-Based Systems}, vol. 237, p. 107852, 2022. [Online]. Available: \url{https://www.researchgate.net/publication/356201584_Deep_learning_for_depression_recognition_with_audiovisual_cues_A_review}
\BIBentrySTDinterwordspacing

\bibitem{AlSahili2024Survey}
\BIBentryALTinterwordspacing
Z.~A. Sahili, P.~Li, G.~Dong, M.~F. Alhamid, M.~Ali, C.~G. L.~G. Cooijmans, M.~M. AbdelMoneim, M.~A. Moni, and K.~K. L., ``Multimodal machine learning in mental health: A survey of data, algorithms, and challenges,'' 2024. [Online]. Available: \url{https://arxiv.org/abs/2407.09020}
\BIBentrySTDinterwordspacing

\bibitem{MultimodalFusionChallenges}
\BIBentryALTinterwordspacing
Y.~Li, S.~Wang, X.~Zhu, and Y.~Zhou, ``Enhancing multimodal depression diagnosis through representation learning and knowledge transfer,'' \emph{BMC Medical Informatics and Decision Making}, vol.~24, no.~1, p.~86, Feb 2024. [Online]. Available: \url{https://pmc.ncbi.nlm.nih.gov/articles/PMC10877283/}
\BIBentrySTDinterwordspacing

\bibitem{Wu2022SurveyHITL}
\BIBentryALTinterwordspacing
M.~Wu, S.~Yuan, and J.~Zhang, ``A survey on human-in-the-loop for machine learning,'' \emph{ACM Computing Surveys}, vol.~55, no.~4, 2022. [Online]. Available: \url{https://doi.org/10.1145/3502260}
\BIBentrySTDinterwordspacing

\bibitem{HITLDefinitionStanford}
\BIBentryALTinterwordspacing
{Stanford HAI}, ``Humans in the loop: The design of interactive ai systems,'' Stanford HAI News, 2020, accessed online. [Online]. Available: \url{https://hai.stanford.edu/news/humans-loop-design-interactive-ai-systems}
\BIBentrySTDinterwordspacing

\bibitem{HACReviewHealth}
\BIBentryALTinterwordspacing
L.~Lai, J.~Wiens, and W.~Weber, ``Human-ai collaboration in healthcare: A review and research agenda,'' \emph{IEEE Journal of Biomedical and Health Informatics}, 2021, accessed via ResearchGate. [Online]. Available: \url{https://www.researchgate.net/publication/349153240_Human-AI_Collaboration_in_Healthcare_A_Review_and_Research_Agenda}
\BIBentrySTDinterwordspacing

\bibitem{HACReviewEducation}
\BIBentryALTinterwordspacing
O.~Zawacki-Richter, V.~I. Marín, M.~Bond, and F.~Gouverneur, ``Human and ai collaboration in the higher education environment: opportunities and concerns,'' \emph{International Journal of Educational Technology in Higher Education}, vol.~21, no.~1, p.~22, 2024. [Online]. Available: \url{https://pmc.ncbi.nlm.nih.gov/articles/PMC11001814/}
\BIBentrySTDinterwordspacing

\bibitem{natarajan2024human}
\BIBentryALTinterwordspacing
S.~Natarajan, S.~Mathur, S.~Sidheekh, W.~Stammer, and K.~Kersting, ``Human-in-the-loop or ai-in-the-loop? automate or collaborate?'' \emph{arXiv preprint arXiv:2412.14232}, 2024. [Online]. Available: \url{https://arxiv.org/abs/2412.14232}
\BIBentrySTDinterwordspacing

\bibitem{AffectiveEthicsRG}
\BIBentryALTinterwordspacing
C.~Reynolds and R.~W. Picard, ``Ethics and information privacy in affective computing,'' in \emph{Proc. Workshop on Affective Computing, CHI 2004}, 2004, accessed via ResearchGate. [Online]. Available: \url{https://www.researchgate.net/publication/215588838_Ethics_and_Information_Privacy_in_Affective_Computing}
\BIBentrySTDinterwordspacing

\bibitem{AffectiveEthicsPDF}
\BIBentryALTinterwordspacing
J.~Gratch, C.~L. Lisetti, and N.~Mavridis, ``Ethical issues in affective computing,'' Chapter in Oxford Handbook of Affective Computing, 2014, accessed online PDF, appears to be a chapter. [Online]. Available: \url{https://people.ict.usc.edu/~gratch/CSCI534/}
\BIBentrySTDinterwordspacing

\bibitem{ExplainableAIHealthcareDeeploy}
\BIBentryALTinterwordspacing
{Deeploy}, ``Explainable ai in personalized mental healthcare,'' Deeploy Blog, 2023, accessed online. [Online]. Available: \url{https://deeploy.ml/explainable-ai-in-personalized-mental-healthcare/}
\BIBentrySTDinterwordspacing

\bibitem{ExplainableAIStressIJRIAS}
\BIBentryALTinterwordspacing
A.~Destiny, ``Leveraging explainable {AI} and multimodal data for stress level prediction in mental health diagnostics,'' \emph{International Journal of Research and Analytical Reviews (IJRIAS)}, pp. 416--425, Jan 2025, published online Jan 15, 2025. Accessed online. [Online]. Available: \url{https://rsisinternational.org/journals/ijrias/articles/}
\BIBentrySTDinterwordspacing

\bibitem{ResponsibleAIEthicsSmythos}
\BIBentryALTinterwordspacing
{Smythos}, ``Establishing ethical guidelines for human-ai collaboration: Best practices and frameworks,'' Smythos Blog/Website, 2023, accessed online. [Online]. Available: \url{https://smythos.com/artificial-intelligence/human-ai-collaboration/}
\BIBentrySTDinterwordspacing

\bibitem{ShneidermanHCAI}
\BIBentryALTinterwordspacing
B.~Shneiderman, ``Human-centered artificial intelligence: Reliable, safe \& trustworthy,'' \emph{International Journal of Human–Computer Interaction}, vol.~36, no.~6, pp. 495--504, 2020. [Online]. Available: \url{https://dl.acm.org/doi/fullHtml/10.1145/3313831.3376275}
\BIBentrySTDinterwordspacing

\bibitem{dauphin2017language}
Y.~N. Dauphin, A.~Fan, M.~Auli, and D.~Grangier, ``Language modeling with gated convolutional networks,'' in \emph{Proceedings of the 34th International Conference on Machine Learning}, ser. Proceedings of Machine Learning Research, vol.~70.\hskip 1em plus 0.5em minus 0.4em\relax PMLR, 2017, pp. 933--941.

\end{thebibliography}
\end{document}